\documentclass[conference]{IEEEtran}

\usepackage{url}
\usepackage{listings}
\usepackage{xcolor}
\usepackage{color}
\usepackage{graphicx}
\usepackage{adjustbox}

\lstdefinestyle{JSON}{
	basicstyle=\ttfamily\footnotesize,
	breakatwhitespace=false,
	breaklines=true,
	captionpos=b,
	keepspaces=true,
	numbers=left,
	numbersep=5pt,
	showspaces=false,
	showstringspaces=false,
	showtabs=false,
	tabsize=2
}

\ifCLASSINFOpdf
\else
\fi

\renewcommand\IEEEkeywordsname{Keywords}

\begin{document}

\newcommand\copyrighttext{	
	\Huge {IEEE Copyright Notice} \\ \\
	\large {Copyright (c) 2017 IEEE \\
		Personal use of this material is permitted. Permission from IEEE must be obtained for all other uses, in any current or future media, including reprinting/republishing this material for advertising or promotional purposes, creating new collective works, for resale or redistribution to servers or lists, or reuse of any copyrighted component of this work in other works.} \\ \\
	
	{\Large Published in:
	2017 European Conference on Networks and Communications
	(EuCNC 2017), June 12-15, 2017, Oulu, Finland} \\ \\ 
	DOI: 10.1109/EuCNC.2017.7980642 \\ \\
	\begin{small}
		Preprint from: \url{https://arxiv.org/abs/1704.08957}\\
		Print at: \url{https://doi.org/10.1109/EuCNC.2017.7980642}
	\end{small}
	
	\vspace{2cm}

	Cite as:\\
	\includegraphics[trim={2cm 23cm 9cm 3.35cm},clip]{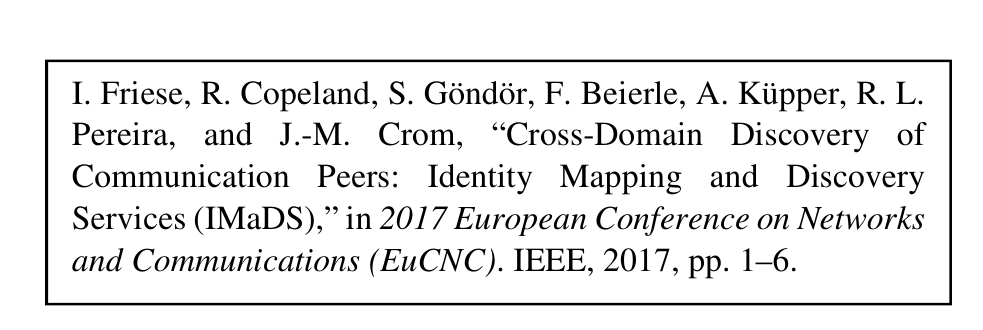}

	\vspace{1.5cm}

	BibTeX:\\ \\
	\includegraphics[trim={2cm 21cm 2cm 3.5cm},clip]{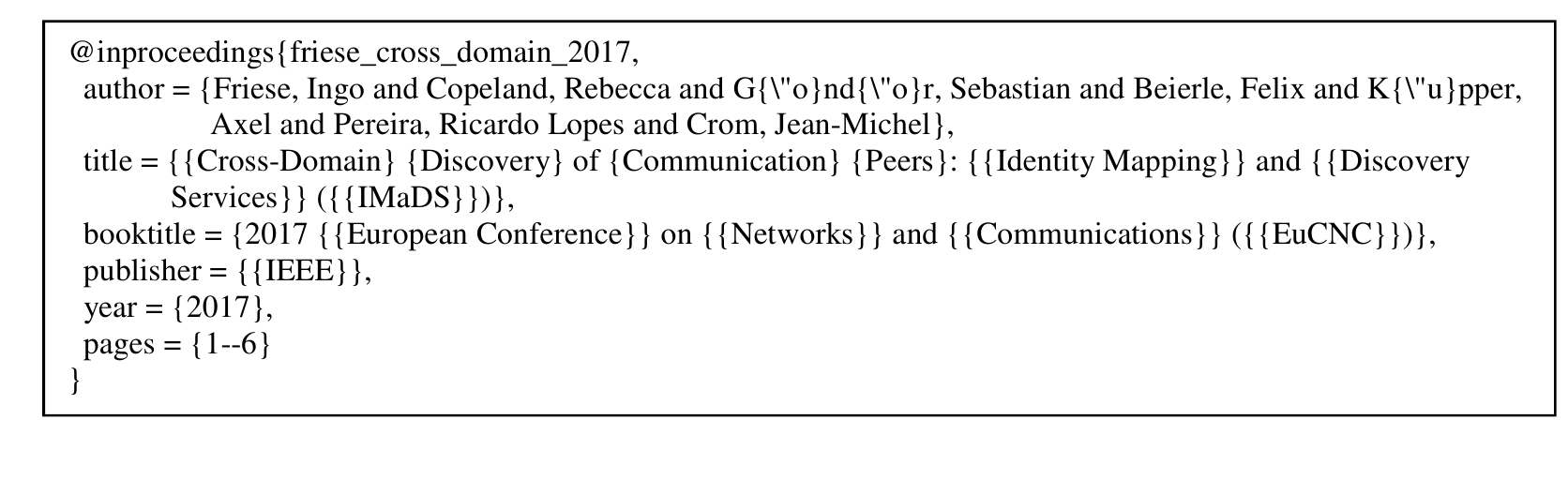}
}
	
\twocolumn[
\begin{@twocolumnfalse}
	\copyrighttext
\end{@twocolumnfalse}
]

\title{Cross-Domain Discovery of Communication Peers\\
\Large{Identity Mapping and Discovery Services (IMaDS)}}

\author{
	\IEEEauthorblockN{
		Ingo Friese\IEEEauthorrefmark{1},
		Rebecca Copeland\IEEEauthorrefmark{2},
		Sebastian G\"{o}nd\"{o}r\IEEEauthorrefmark{3},
		Felix Beierle\IEEEauthorrefmark{3},
		\\Axel K\"{u}pper\IEEEauthorrefmark{3},
		Ricardo Lopes Pereira\IEEEauthorrefmark{4},
		Jean-Michel Crom\IEEEauthorrefmark{5}}
	\IEEEauthorblockA{
		\IEEEauthorrefmark{1}Telekom Innovation Laboratories, Deutsche Telekom AG, Berlin, Germany\\
		Ingo.Friese@telekom.de}
	\IEEEauthorblockA{	
		\IEEEauthorrefmark{2}Institut Mines Telecom, Telecom SudParis, Paris, France\\
		rebecca.copeland.uk@gmail.com}
	\IEEEauthorblockA{	
		\IEEEauthorrefmark{3}Service-centric Networking, Technische Universit\"{a}t Berlin, Berlin, Germany\\
		\{sebastian.goendoer, beierle, axel.kuepper\}@tu-berlin.de}
	\IEEEauthorblockA{	
		\IEEEauthorrefmark{4}INESC-ID / IST, University of Lisbon, Lisbon, Portugal\\
		ricardo.pereira@inesc-id.pt}
	\IEEEauthorblockA{
		\IEEEauthorrefmark{5}Orange Labs, Rennes, France\\
		jeanmichel.crom@orange.com}
}

\maketitle

\begin{abstract}
The upcoming WebRTC-based browser-to-browser communication services present new challenges for user discovery in peer-to-peer mode. Even more so, if we wish to enable different web communication services to interact. This paper presents Identity Mapping and Discovery Service (IMaDS), a global, scalable, service independent discovery service that enables users of web-based peer-to-peer applications to discover other users whom to communicate with. It also provides reachability and presence information. For that, user identities need to be mapped to any compatible service identity as well as to a globally unique, service-independent identity. This mapping and discovery process is suitable for multiple identifier formats and personal identifying properties, but it supports user-determined privacy options. IMaDS operates across different service domains dynamically, using context information. Users and devices have profiles containing context and other specific information that can be discovered by a search engine. The search results reveal the user's allocated globally unique identifier (GUID), which is then resolved to a list of the user's service domains identities, using a DHT-based directory service. Service-specific directories allow tracking of active endpoints, where users are currently logged on and can be contacted.
\end{abstract}


\pubid{978--1--5386--3873--6/17/\$31.00~\copyright~2017 IEEE}

\section{Introduction}

\begin{figure}[!b]
	\begin{minipage}{\columnwidth}
		978-1-5386-3873-6/17/\$31.00 \copyright 2017 IEEE
	\end{minipage}
\end{figure}

The advent of WebRTC \cite{bergkvist2012webrtc} is poised to change the landscape of web calling, moving from specialized Internet VoIP (Voice over IP) services to generic browser-based calling from any visited website. While currently VoIP service providers do not interwork, often by choice (to encourage viral growth), browser-based web calling would seek cross-browser interactions, to enhance website businesses who regard communications as a means to an end. Such web calling 'tools' must overcome issues not faced by current communication services. WebRTC has simplified peer media exchange over the Internet, but there are still major inhibitors for inter-service VoIP compatibility. The work described in this paper addresses three of these: common identity formats or authentication, discovery of identity partners, and discovery of availability status.

There are no developed standards to enable web services to interpret generic user identifier formats, such as the international telephone numbering scheme. Where necessary, these traditional numbers are the only globally recognizable format. 
Additionally, unlike mobile networks, web calling services need to be aware of the presence status of users, including information about the used service provider and support means of alerting them to incoming communication.
Since web services can be accessed from different devices, there is another problem of finding users, since the procedure is anchored at a particular endpoint.

The reTHINK project\footnote{reTHINK project website: \url{http://rethink-project.eu}} developed an Identity Mapping and Discovery Service (IMaDS) in order to find contactable and active endpoints (software that handles interactions within terminals/devices). This service has been developed for WebRTC peer-to-peer (P2P) calling, using specially designed reTHINK endpoint clients, but IMaDS scope may be applied to other services, regardless of the protocol or the type of user.
The mapping and availability discovery mechanisms can apply to context-aware applications that require dynamic intervention and timely reactions. The IMaDS approach is pluralistic and open. It is designed to encourage interworking between service providers who offer a variety of communication services between users, where users may be humans or machines. IMaDS protects the privacy of its users and provides a fine grained access configuration. It can be used across domains and different identifier formats. It enables searching for endpoints within a specific context, which is useful in Smart City and Internet of Things domains, and many more.
While traditional VoIP servers can provide 'presence' information for contactable destination parties within a single service, connecting peering users on different services presents a far more difficult task. Establishing users' ability to receive communication is critical, where there is no central server to log the attempt. Not only the user identifier needs to be discovered, but also which device and which service are currently used. 
A browser-based communication client at the caller's device needs to know the dynamically changing address of the communication client at the called endpoint in order to connect the media (messages or voice/video) between the browsers (using the browser's WebRTC APIs), device to device. Users may be logged in to different service domains and service providers, and may use dynamic environments where clients are 'on' or 'off' frequently. Unlike calling via a network server, P2P interactions require both parties to be 'on', or the communication attempt would fail unconditionally, with no trace of the attempt.
The reTHINK project has developed P2P communication endpoints named hyperties (hyper-entities) \cite{crom2016implementation}. They are micro-services that are downloaded onto the device runtime environment. There are different hyperties for different services, e.g., communication for classic voice, video, and chat services, or inter-object interactions for Internet of Things devices. A hyperty is typically implemented as a piece of JavaScript code from a library on a service provider's website, which is downloaded to the user's browser. The hyperty utilizes the browser's built-in WebRTC APIs, without the need for plugins. Session setup is also simplified and all that is required is the URL of the communication partner. IMaDS is able to retrieve this URL and ascertain the user availability on a particular device and its current IP address.

The rest of the paper includes the state-of-the-art in Section II, the IMaDS approach and methods in Section III, the usage scenarios in Section IV, and conclusions in Section~V.

\section{Related Work}

The WebRTC standard draft \cite{bergkvist2012webrtc} discusses identities and authentication, but does not specify how to find communication endpoints. Currently, WebRTC session setup is not defined in standards. The IETF draft in \cite{copeland2016requirements} lists requirements for identity discovery considering different privacy levels.
The standard directory service is provided by Lightweight Directory Access Protocol (LDAP) \cite{zeilenga2006lightweight}, which allows organizing information in 'Directory Information Trees'. However, this is too restrictive for an open system where there is no “one size fits all” schema, no single identifier format, and variable context information.
Discovery methods fall into two approaches: distributed peer-to-peer (P2P) discovery mechanisms and central directory systems. In P2P discovery, peers are assumed to know only their neighbors, but every discovery query is routed to other peers.
The Gnutella protocol \cite{clip2007gnutella} is a classic example of this, where queries are flooded. Peer discovery is fault-tolerant and works even when some peers are offline, but it produces high network loading and does not scale.
BitTorrent uses a central directory, the tracker, to enable peers to find other peers~\cite{cohen2003incentives}. Clients register when they log in, and their status is changed when they sign off.
The tracker records the service, device, and IP address, so that clients are contactable and discoverable. The central directory approach has a single point of failure, but the network traffic is minimal and easily manageable.
A central directory is also used in communication services like Google Hangout\footnote{Google Hangouts: \url{https://hangouts.google.com}} and Skype\footnote{Skype: \url{https://skype.com}}.
However, this approach allows only a single service to discover its own users, and there is no possible discovery of endpoints that are registered with another service. What is required is a collaborative ecosystem of such central directories that allow searching across participating services.

The Domain Name Service (DNS) \cite{mockapetris1987domain} is, in fact, a well-established discovery service. It maps domain names to IP addresses of the actual endpoints. The DNS queries are served by a local resolver on the DNS server in the same domain. If the address cannot be resolved locally, the query is resolved by iteratively contacting the authoritative servers for the domain in a hierarchical way.
The DNS is well governed, but its reliance on extensive caching results in delays in the propagation of new entries or updates. 
To enhance performance, DynDNS \cite{bound1997dynamic} dynamically updates domain names. Papas et al. \cite{pappas2006comparative} have also proposed architecture optimizations to increase DNS performance. DNS has no built-in privacy mechanisms and it has format restrictions. ENUM \cite{hoeneisen2011iana} provides a mapping scheme to the telephone numbering standards (ITU E.164 international numbering plan for public telephone systems), but not to alternative identifiers, so it cannot deal with cross-service identity mapping.

Distributed Hash Tables (DHT) are decentralized and heavily distributed systems that allow lookup services to query data using key/value pairs. While DHTs allow to lookup data as fast as $O(log(n))$, writing performance strongly depends on individual nodes \cite{maymounkov2002kademlia}. Also, DHT security is a serious concern, as described in \cite{urdaneta2011survey}, which concludes that a combination of different methods would be effective but not perfect. Most DHT deployments depend on Kademlia's data replication and redundant routing.
Replication, essential for ensuring the required robustness despite peer churn, makes data modification expensive, as data has to be synchronized to several nodes.

\section{Identity Mapping and Discovery Process}
\begin{figure*}
	\centering
	{
		\includegraphics[trim={0 219 0 65},clip,width=1.0\textwidth]{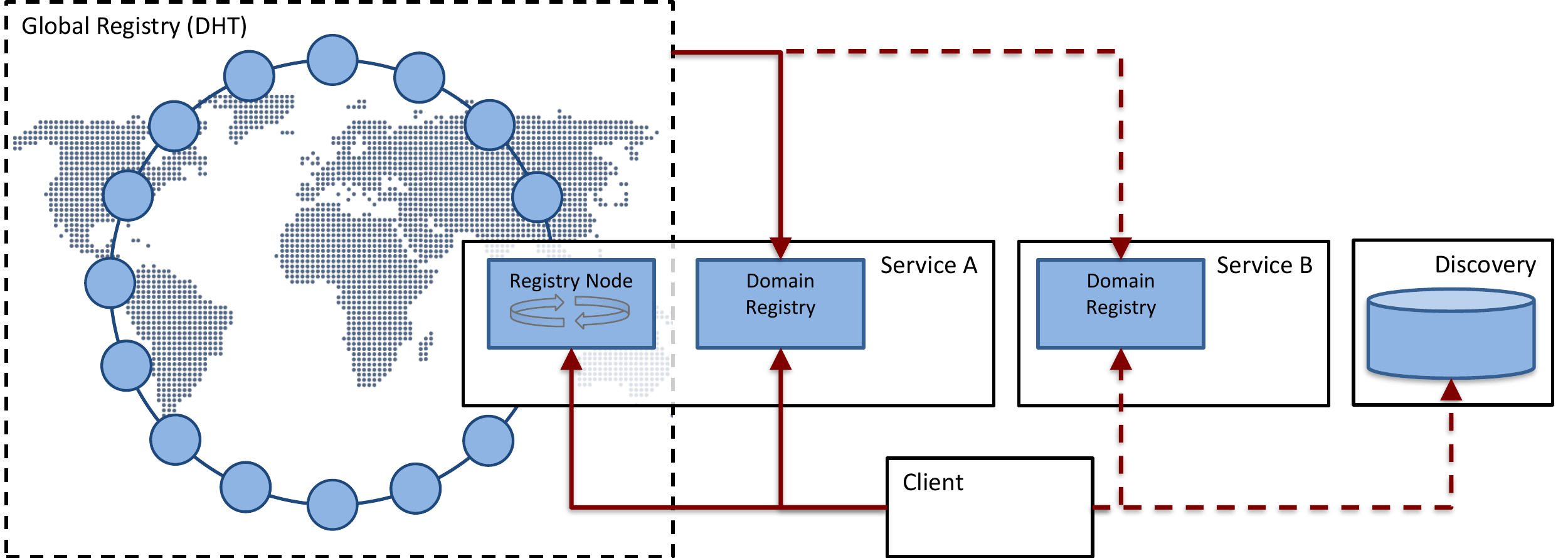}
		\caption{Interplay of IMaDS components}
		\label{architecture}
	}
\end{figure*}

IMaDS was designed to provide a mapping between users and the endpoints where the services they choose to use are available, while maintaining their privacy.
It should decouple identity from service provider, thus ensuring users are free to change service providers without being locked in.
It should also have the ability to evolve search beyond what is currently envisioned, e.g., by combining service and endpoint data with social data and other sources of information such as user profiles.
IMaDS accomplishes this, while also fulfilling a set of non-functional requirements: be scalable for worldwide use, provide low-delay reads, be highly available, prevent identity theft. 
IMaDS requirements go beyond the capabilities of traditional LDAP and DNS, so a fresh approach is needed to provide greater versatility and flexibility. To facilitate cross-domain discovery of peers in a fast and reliable fashion, IMaDS encompasses three major functions that can be used independently (see Figure \ref{architecture}):

\begin{itemize} \itemsep2pt \parskip0pt \parsep0pt
	\item{\textbf{Domain-Registry Queries:}} A classic directory service within the domain of one service provider. It provides a list of URLs for currently running hyperties (communication endpoints) for a given username in a certain domain. Each service provider makes available its own Domain Registry.
	\item{\textbf{Global Registry Queries:}} A DHT-based global directory service, mapping user identifiers to the associated service domains. The Global User ID (GUID) is resolved to a digitally signed dataset, comprising all associated service domains and respective identities of the user account. 
	\item{\textbf{Discovery Service:}} A dedicated directory combined with a search engine and a privacy policy database. A user can create a profile that contains text, description, hashtags, and geographical information, linked to the GUID. 
\end{itemize}

\begin{figure*}[t]
	\centering
	{
		\includegraphics[trim={113 266 139 80},clip,width=0.86\textwidth]{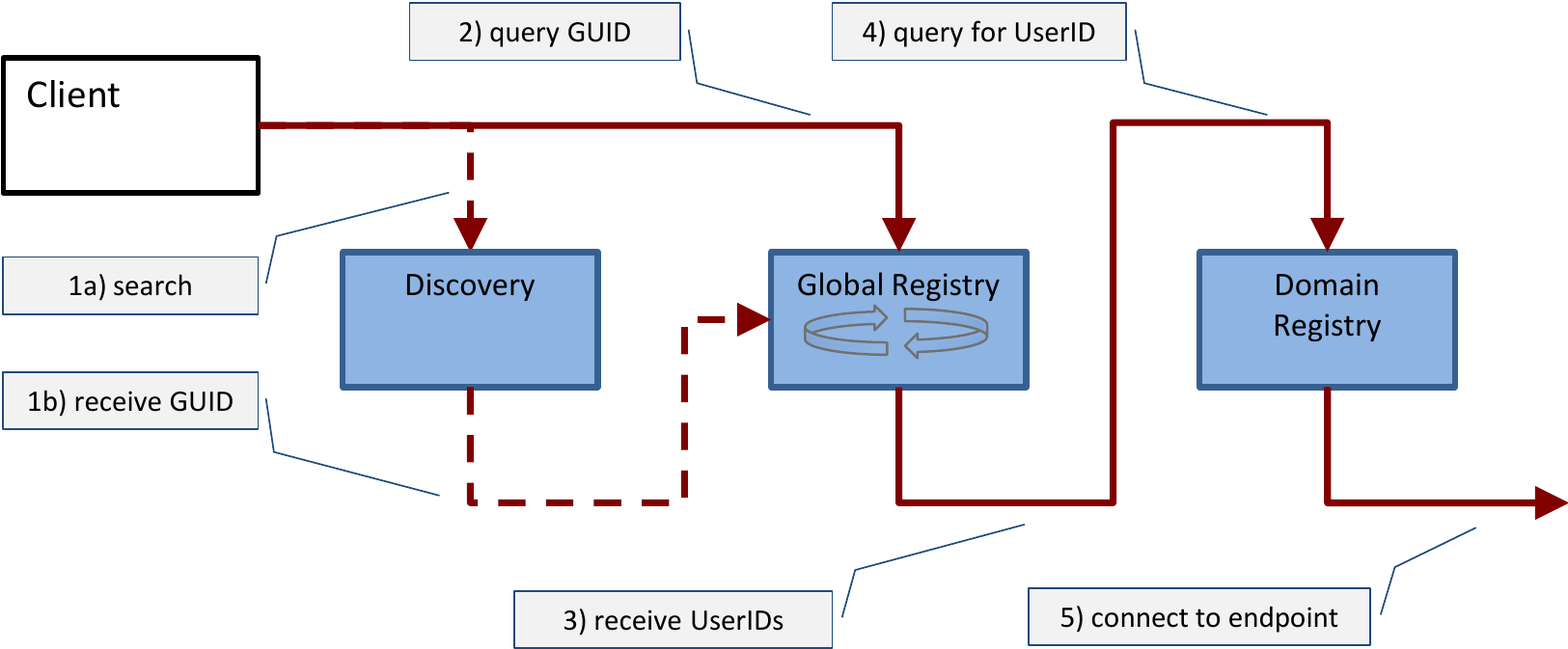}
		\caption{IMaDS process of discovering and resolving identifiers}
		\label{resolving}
	}
\end{figure*}

Figure \ref{resolving} shows the interworking of all three parts of the mapping and discovery process starting from an intuitive search query in a natural language, to mapping a unique GUID to a temporarily available URL of a client that can be connected to.

\subsection{Domain Registry}
The Domain Registry's main function is to provide a mapping from a userID to the set of endpoints available on that user's devices.
It is assumed that every service provider will run its own instance of the Domain Registry that will store the endpoints for the hyperties provided by that service provider.
When an application is run by the user on one of its devices, the application will instantiate the hyperties that provide the services that the application requires.
In the reTHINK framework, the runtime will automatically register each hyperty instance with the service provider's Domain Registry.
The Domain Registry uses soft state, the information about the hyperty instance must be periodically refreshed or it will be removed.

The Domain Registry provides a REST API that uses the userID and Service Provider domain as the first hierarchical level of the URLs.
Under each userID, hyperty instances are stored, one for each service running on each device.
For each hyperty instance, several pieces of information are used, such as the URL that should be used to reach the endpoint and its supported capabilities (e.g., voice, video, or chat).
Applications query the Domain Registry in order to determine how to contact a user.
Typical queries include asking for all of a user's available endpoints that provide a certain capacity (e.g., video calling) or asking for the current availability and reachability status of a previously known endpoint.
Endpoints may became available or unavailable as the device loses Internet connectivity or the user chooses not to be available.
In reTHINK, applications can request to be notified when a currently unavailable endpoint comes back online.

Performance and availability are critical aspects of the Domain Registry, as it is in the critical path for call setup.
Due to network mobility and device churn, the Domain Registry is write intensive, and consistency is required.
On the other hand, each instance only provides service to a single service provider.
The use of a REST API enables the use of well known load-balancing and failover techniques.
Our deployment architecture uses replicated load balancers that distribute requests to a set of application servers.
The application servers access a Cassandra Database, distributed and replicated over several nodes.
No single point of failure exists.

\subsection{Global Registry}

The Global Registry is based on the idea of the Global Social Lookup Service \cite{goendoer2016distributed} and is designed for collaboration between different domains, which allows them to discover destination parties that are not within their own domain. While each Domain Registry remains an internal facility of a service provider, the Global Registry is organized as a domain-independent service that allows other services to query it.
To allow cross-domain lookup of identifiers, the Global Registry has been designed as a component separated from the Domain Registry, built on P2P-technology to distribute control over the system using a Kademlia-based DHT \cite{maymounkov2002kademlia}. 
This way, both database and workload are distributed over a high number of servers, which organize communication and management of the network automatically. The fault tolerant architecture of the Kademlia architecture minimizes the probability of data loss due to indirect replication of all data to multiple nodes, as well as overall system stability due to self-management of the nodes.
As users don't subscribe to new services that often, the Global Registry's load is expected to be heavily read biased.
This is the type of load best served by a DHT, which is scalable to provide a worldwide service.
Also, absolute consistency is not required.
Thus eventual small delays in the propagation and replication of information are acceptable.
 
Nodes of the DHT are run by service operators, who cannot exert control over the managed data or the functionality of the system altogether. This not only achieves separation of concerns, but also prevents the system from being vulnerable due to not exposing a single point of failure. Furthermore, concerns of allowing a single operator to control the overarching cross-domain lookup altogether are avoided.
The Global Registry service is implemented as a decentralized directory of users, where users can be enrolled with one or with multiple service providers at the same time.
To associate users with different services, a service-independent identifier is necessary. WebRTC architecture introduces the notion of a service-independent identity that is managed by an Identity Provider (IdP) and can provide authentication to any service. Such identities may be authenticated directly in a P2P fashion \cite{copeland2016requirements} for WebRTC calling, but they also need to be associated with the application based userIDs, to be recognized by their applications.

Users are identifiable via a self-asserted, domain agnostic identifier, the GUID, which is linked with the domain-specific userIDs, according to user instructions. This way, there is a stable, unifying single identifier that can be used even when the identity providers themselves are changed. The GUID is a long alphanumeric identifier that is generated by a key derivation algorithm for creating random and unique keys. 
The result is a typically non-human readable identifier with a suitable amount of randomness to guarantee global uniqueness. An example GUID is \texttt{OuLbxRKrYcyvfU8cmSqdyxoBq7j3IAypd20iuPDsIg}. Such identifiers may be provided to others by the users, but the main purpose of such unmemorable identifiers is to link the user to multiple service identifiers via the application facilities. The client uses the provided GUID to discover a list of service providers and contactable userIDs. A user may have any number of GUIDs and link to them to a different range of services, if so wished.
As the GUID is a self-asserted identifier, the users' datasets are managed by the users themselves using an IMaDS compatible client application for key management and lookup of communication peers. To create the GUID, a user creates an ECDSA public key pair \cite{johnson2001elliptic} and hashes its public key with a cryptographic salt of fixed length using the key derivation function PBKDF\#2 \cite{kaliski2000pkcs} and SHA256. The binary output of 256 bit length is encoded using Base64URL
\cite{josefson2006base} and then used as the GUID.

To allow resolving a GUID to service URLs and userIDs, the GUID, public key, salt, userIDs, as well as additional information are stored in a JSON-based dataset. This dataset is then published as a signed JSON Web Token (JWT) \cite{jones2015json}, where the digital signature is created using the private key of the user's key pair. The dataset is then published in the Global Registry using the service's RESTful API for read and write operations, where the GUID is used as the lookup key. Thus, any of the collaborative services who share the Global Registry can resolve a known GUID to the respective user's dataset, while the integrity of the dataset's contents is guaranteed through the digital signature of the JWT.
As GUIDs are generated in a distributed fashion, attacks are possible that aim at taking over an IMaDS identity \cite{goendoer2016distributed}.
Since GUIDs are derived directly from the users public key, an attacker would need to create a valid signature for a forged dataset.
Simple replacement of the key itself is not possible, as this would inevitably also change the GUID, resulting in a deflected attack.
Hence, for a successful attack aimed at taking over an identity, an attacker would need to create a key pair and salt that creates a collision in PBKDF\#2 with SHA256.
According to the birthday bound, for a $1\%$ chance of a collision, $4.8 \times 10^{37}$ key pairs and salts would be required, making a successful attack unreasonably hard to achieve. 

\subsection{The Discovery Service}

The Discovery Service is able to utilize any information that can be queried in a search engine. Thus, active endpoints for a user can be found even without knowing the service provider, the service-based identity, or even the GUID. The basic idea of IMaDS discovery is to find users in a natural and intuitive way, like any natural language search engine. This way, users can build search queries using facts describing the search queries' target, for example, "Alice," "Deutsche Telekom," and "Berlin":
\vspace{3pt}
\lstset{
	basicstyle=\ttfamily\footnotesize, 
	breakatwhitespace=false,           
	breaklines=true,                   
	captionpos=b,
	backgroundcolor=\color{white},     
}

\begin{lstlisting}[frame=single]
https://rethink.tlabscloud.com/discovery/rest/discover/lookup?searchquery=Alice+Deutsche+Telekom+Berlin
\end{lstlisting}

If the search is successful, the discovery service will provide one or several profiles matching the search query.

\begin{lstlisting}[frame=single]
{
  "instanceID":"telekom1",
  "responseCode":201,
  "searchString":"Alice",
  "results":[{
    "resultNo":0,
    "hashtags":"#reTHINK #Telekom",
    "description":"My profile”, 
    "GUID":"WabRS8ZRswDNUIYtqF-j0nHQZmQVRLJimvqIGIYMz50",
    "headline":"Testprofile Alice",
    "contacts":"www.telekom.de",
    "hasGUID":"true",
    "hyperties": [{
      "url":"hyperty=hyperty%3A%2F%2Frethink.tlabscloud.com%2F4246e263-eb54-4a",
      "userID":"uid=user://gmail.com/alice",
      "media":"VIDEO",
      "provider":"Deutsche Telekom"
    }]
  }]
}
\end{lstlisting}

A user who wants to be found can create an account at the IMaDS discovery service, with any number of profiles. Alice, for example, might have one profile for her business life and another one for private use. The user can choose what profile information is stored, e.g., employer, personal address etc. Hashtags can be added if the user wants to be found under certain keywords or interests, for example, \#football, \#tango, or \#paris. Furthermore, a user can specify a number of contacts, such as email addresses or phone numbers. Additional search fields, such as a profile photo, may be added to the index of the search engine.
Thus, every profile can be connected to a particular context by adding keywords, hashtags, or even geographical information. When a valid GUID is linked to this profile, it is dynamically mapped via the global and domain registries to the endpoint activities. 
Web search engines find most of its content through 'web-crawling.' Software agents crawl the whole web by calling every hyperlink and indexing every website. As a result, the owner of the website has no control when his content gets visible and when it will be deleted. In IMaDS, users are fully in control of their information. The data is published when a profile is created and unpublished whenever a profile is deleted. Besides this 'all or nothing' approach, IMaDS also enables different levels of data visibility. For every profile, a visibility option can be chosen. A user profile can be visible for all users, for IMaDS users only, or just for favorite users. This way, a profile can be configured to be visible by a small group of known users, for example, friends, class mates, or colleagues.

The service supports advanced privacy and service features. Privacy is enabled through changing endpoints: the service can be configured for a communication client to get a new URL with every restart or after a configurable time period. Thus, a malicious party cannot contact this client again, even if the URL is cached or leaked. When a user wants to contact the same party, a fresh discovery process is required.
The discovery service enables:

\begin{itemize} \itemsep2pt \parskip0pt \parsep0pt
	\item{\textbf{Privacy through configurable visibility:}} Users can configure the visibility of their clients or devices to other users. IMaDS enables setting fine grained policies, defining who has access to what information. The user is in full control of information that is shown to others.
	\item{\textbf{Multi-domain, multi-Identifier:}} The service supports discovery across different service domains and service providers. The process works for several identifier formats, and can handle URLs, phone numbers, postal addresses, etc.
	\item{\textbf{Context Sensitivity:}} IMaDS searching factors can be linked to any context descriptor, such as user roles, geo-location data, or keywords. This enables the IMaDS functions to become basic components for many other services that require dynamic discovery of actors (users or things) in volatile environments.
\end{itemize}

\section{IMaDS Usage Scenarios}

A typical usage scenario of the IMaDS services comprise a user Alice, who wants to connect to another user Bob, who is registered in one or more service domains. If required, Alice would query the Discovery component to search for Bob using facts that describe Bob. By resolving Bob's GUID via the Global Registry, she receives a list of service domains Bob is registered with. As each userID in the returned dataset also comprises information about the respective service provider's Domain Registry, Alice can now query the Domain Registry for Bob's communication endpoint and establish a connection.

\subsection{Alternative Services}

The IMaDS functions were created to find WebRTC peering parties, but can be used in many other scenarios, as shown in the following examples:

\begin{itemize} \itemsep2pt \parskip0pt \parsep0pt
	\item{\textbf{Ad Hoc networks:}} The Domain Registry can handle clients that popup and disappear within a short time period.
	\item{\textbf{Heterogeneous networks like IoT:}} The process is protocol-agnostic. Discovery and also Global Registry could map a GUID to any protocol, not only to HTTP.
	\item{\textbf{Highly distributed infrastructures:}} The separation of Domain and Global Registry as well as the design of Global Registry as a DHT allows for a high level of distribution and scalability.
	\item{\textbf{Highly privacy-aware services:}} The discovery process allows configuring visibilities, ranging from being visible for closed user groups to being visible for the whole World Wide Web.
	\item{\textbf{Community Services:}} The ability to tag endpoints in profiles with context, keywords, and geo data enables a plethora of services. IMaDS is able to find users with common interests or from the same city or area that can be provided with certain information.
\end{itemize}

\subsection{Privacy-Preserving Context-Based Services}

The discovery mechanism for activated micro-services (such as reTHINK hyperties) can be exploited for other types of activities, under the same infrastructure. Instead of discovering users' availability to receive calls, the discovery requests will return current users' active usage of a particular function, on a particular device, thus discovering fine-tuned contexts dynamically. This enables observers to know not only when a user is online or logged in to a service, but when the user is actively engaged in a particularly monitored activity. This also means that the user's attention is focused on that service, at that moment, on that device, and at a specific location. This intelligence enables service providers to deliver highly targeted support, additional information, advice, or any other type of time-critical context-based functions. Unlike other such services, IMaDS users can control this effect by switching it on or off via the privacy capability for each associated service, so that the privacy rules are selective.

\subsection{Future Development}

Currently, the discovery part containing the IMaDS search engine is a single instance. In the next version, one discovery instance can exchange queries and results with others. Thus, different providers can build their repository of profiles with their own user interface and own type of information.
The reTHINK project will develop new services that can be modular add-ons for IMaDS:

\begin{itemize} \itemsep2pt \parskip0pt \parsep0pt
	\item \textbf{Trust-Engine:} This is an engine \cite{crom2016implementation} to evaluate the level of trust based on past transactions. IMaDS could take trust-level into account, and mark trusted communication endpoints accordingly.
	\item \textbf{Social graph:} The social graph \cite{BGK2015} stores interactions and relationships between peers. This could also be used as an optional parameter for a discovery query.
	\item \textbf{Usability:} The adoption of IMaDS depends on its usability for both developers and users. In the next project phase, reTHINK will focus on a consistent and secure, yet easy-to-use registration process for service providers and users.
\end{itemize}

\section{Conclusion}

IMaDS was developed within the reTHINK project to find communication endpoint micro-services (hyperties). Features like privacy configurations, multi-domain support, and context sensitivity make this service a good candidate to be extended to other services. IMaDS introduces two main new functions: i) Mapping user identities of multiple services to a user's chosen independent identifier; the IMaDS mapping of identities links domain-based service identities and a service-independent GUID, so that users can control their identity associations and assign different privacy rules, if so wished; ii) Discovery of 'live' micro-services (such as the described reTHINK hyperties, or other critical monitored processes that require downloading of micro-service code) when the user activate them or switches them off. 
The IMaDS discovery service is inclusive, not service-related, and responds with current contact information for the searched user, with several options of both services and devices that can be reached via peer-to-peer interaction. This approach is based on peering mechanisms and user self-management of their identity structures. The novel method registers activation of micro-services (hyperties) in local Domain Registries and searches the Global Registry of collaborating services to reveal currently available users' contacts in any domain, without the need for a central authority. 
Although this paper describes concepts, methods, and design for WebRTC communication as developed in the reTHINK project, the IMaDS functions are proven very powerful, and can be used in many more service solutions, such as monitoring critical micro-services usage and IoT device interactions.

\section*{Acknowledgment}

This work has received funding from the European Union's Horizon 2020 research and innovation program under grant no. 645342, project reTHINK.

\bibliographystyle{IEEEtran}

\end{document}